\documentclass[twocolumn,secnumarabic,amssymb, nobibnotes, aps, prl, superscriptaddress]{revtex4-2}

\usepackage[]{graphicx}
\usepackage{tabularx}
\usepackage[usenames,dvipsnames]{color}
\usepackage{soul}
\usepackage{ulem}
\usepackage{bm}
\usepackage{amsmath}
\usepackage{lineno}
\usepackage{gensymb}

\usepackage{times}
\usepackage{amsfonts}
\usepackage{mathrsfs}
\usepackage{graphicx}% Include figure files
\usepackage{dcolumn}% Align table columns on decimal point
\usepackage{bm}% bold math
\usepackage{color}

\usepackage[colorlinks,bookmarks=false,citecolor=blue,linkcolor=red,urlcolor=blue]{hyperref}
\usepackage{multirow}
\usepackage{physics}

\begin{document}

%\title{dc and double-loop dc SQUID based on a multi-component superconductor with broken time-reversal symmetry as a superconducting diode}

%\title{Superconducting diode effect in a dc SQUID based on a multi-component superconductor as a probe for identifying the broken time-reversal symmetry}

\title{Supercurrent Diode Effect in Josephson Interferometers with Multiband Superconductors}

%\title{Possible probe for identifying the broken time-reversal symmetry in a multi-component superconductor using superconducting diode effect in a dc SQUID}

\author{Yuriy Yerin}
\affiliation{CNR-SPIN, via del Fosso del Cavaliere, 100, 00133 Roma, Italy}
\author{Stefan-Ludwig Drechsler}
\affiliation{Institute for Theoretical Solid State Physics, Leibniz-Institut für Festkörper- und Werkstoffforschung IFW-Dresden, D-01169 Dresden, Helmholtzstraße 20}
\author{A. A. Varlamov}
\affiliation{CNR-SPIN, via del Fosso del Cavaliere, 100, 00133 Roma, Italy}
\affiliation{Istituto Lombardo ``Accademia di Scienze e Lettere'', via Borgonuovo,
25 - 20121 Milan, Italy}
\author{Francesco Giazotto}
\affiliation{NEST Istituto Nanoscienze-CNR and Scuola Normale Superiore, I-56127, Pisa, Italy}
\author{Mario Cuoco}
\affiliation{CNR-SPIN, c/o Universit\'a di Salerno, I-84084 Fisciano (SA), Italy}

\date{\today}

\begin{abstract}
%We investigate nonreciprocal supercurrent effects in superconducting quantum interferometric devices (SQUID) based on Josephson junctions with single and multiband order parameters.  We show that the magnetic field is able to modulate the amplitude and sign of the supercurrent rectification independently of the multiband character of the employed superconductors. We investigate the role of zero and antiphase pairing among different bands in setting out nonreciprocal effects demonstrating that the rectification is not sensitive to the $\pi$-pairing. The inclusion of multiband superconductors that break time reversal symmetry leads to a remarkable signature in the rectification: the average of the rectification amplitude over multiples of quantum fluxes is not vanishing.  

We investigate nonreciprocal supercurrent phenomena in superconducting quantum interference devices (SQUIDs) that integrate Josephson junctions with single and multiband order parameters, which may exhibit time-reversal symmetry breaking. Our results show that the magnetic field can independently control both the amplitude and direction of supercurrent rectification, depending on the multiband characteristics of the superconductors involved. 
We analyze the effects of zero and antiphase ($\pi$) pairing among different bands on the development of nonreciprocal effects and find that the rectification is not influenced by $\pi$-pairing. Furthermore, we demonstrate that incorporating multiband superconductors that break time-reversal symmetry produces significant signatures in rectification. The rectification exhibits an even parity dependence on the magnetic field and the average rectification amplitude across quantum flux multiples does not equal zero. These findings indicate that magnetic flux pumping can be accomplished with time-reversal symmetry broken multiband superconductors by adjusting the magnetic field. 
%which results from the induced net nonreciprocal supercurrent. 
Overall, our findings provide valuable insights for identifying and utilizing phases with broken time-reversal symmetry in multiband superconductors.

%We present a proposal to exploit the diode effect in a dc SQUID based on Josephson junctions between single-band and multi-band superconductors as a probe for the detection of the state with the broken time-reversal symmetry. We show that if a state with broken time-reversal symmetry emerges in a multi-band (multicomponent) superconductor, the superconducting diode effect possesses a number of unique features that distinguish it from a similar effect caused by the asymmetry of the properties of Josephson junctions in a dc SQUID.  
\end{abstract}
%\pacs{}
\maketitle

%\section{Introduction}

\noindent\large{\textbf{Introduction}}\normalsize\\ 
There is an exciting and rapidly evolving field of research in nonreciprocal superconductivity. Nonreciprocal superconductivity refers to a phenomenon where the properties of a supercurrent change based on the direction in which it flows. This behavior is distinct from conventional superconducting currents, which typically exhibit symmetric flow.
Studying nonreciprocal effects and the rectification of supercurrents not only enhances our understanding of fundamental physics but also paves the way for potential applications in next-generation electronic devices and quantum technologies. \cite{upa24,cas24,ing24,pag24}. 
%The complexity and richness of superconducting materials and their interactions, which could lead to transformative advancements in both theoretical understanding and practical applications.

Several directions have been explored in the design of nonreciprocal superconducting effects, with a particular emphasis on the role of materials in developing supercurrent rectifiers. Research has shown nonreciprocal supercurrent effects in both traditional superconducting materials and a wide variety of other materials. These include non-centrosymmetric superconductors \cite{WakatsukiSciAdv2017, and20, Zhang2020}, two-dimensional electron gases, polar semiconductors \cite{Itahashi20}, patterned superconductors \cite{Lyu2021}, superconductor-magnet hybrids \cite{wu22, Narita2022, Sun_altermagnet}, Josephson junctions that incorporate magnetic atoms \cite{tra23}, twisted graphene systems \cite{lin22}, and high-temperature superconductors \cite{Ghosh2024}.

Numerous intrinsic and extrinsic mechanisms have been put forward to develop a supercurrent diode capable of controlling the rectification in both sign and amplitude. In this context, it is widely acknowledged that breaking time-reversal and inversion symmetries constitutes a fundamental requirement for realizing a nonreciprocal supercurrent. Various physical scenarios and mechanisms have been explored to achieve nonreciprocal supercurrent effects, including the manipulation of Cooper pair momenta \cite{lin22,pal22,yuan22}, the exploitation of helical phases \cite{Edelstein95,Ilic22,dai22,he22,Turini22}, magnetic texture and magnetization gradients \cite{roig2024}, the presence of screening currents \cite{hou23,sun23}, and supercurrents associated with self-induced fields \cite{kras97,GolodNatComms2022}.

Typically, the breaking of time-reversal symmetry is accomplished by applying external magnetic fields. Additionally, the presence of vortices is anticipated to contribute to supercurrent diode effects, and this aspect has been investigated across various physical configurations \cite{GolodNatComms2022,sur22,GutfreundNatComms2023,Gillijns07,Ji21,He19,MarginedaCommunPhys2023,Paolucci23,Greco23,Lustikova2018,Itahashi20, Debika2024, Mao_spin, Cheng_dot, Sun_magn_imp, Sun_altermagnet, Scheurer1, Scheurer2, Soori_2025,Suri2022}.
Moreover, innovative proposals for magnetic field-free superconducting diodes have emerged, utilizing magnetic materials incorporated into specifically designed heterostructures \cite{wu22,Narita2022}. Alternatively, mechanisms involving back-action supercurrents effects due to applied gating \cite{Margineda2023} offer a pathway to achieve superconducting rectification effects without relying on external magnetic fields or magnetic materials. This diverse range of approaches highlights the ongoing advancements in superconducting diodes, paving the way for novel applications and technologies.

Regarding the use of sources of time-reversal symmetry breaking, a large variety of superconducting materials display characteristics indicative of spontaneous time-reversal symmetry breaking (TRSB) below their transition temperatures \cite{Kallin_2016,Wyso2019,Ghosh_2021,Poccia}. 
This phenomenon can occur in the presence of spin or orbital angular momentum of the Cooper pairs, due to disorder \cite{Stanev,Corticelli,Lee2009,Maiti2015}, or in the presence of $\pi$-pairing—characterized by an antiphase relationship between the order parameters across different bands. $\pi$-pairing in multiband electronic structures frequently occurs together and can signal the presence of unconventional superconducting phases. This is evident in various materials platforms such as iron-based superconductors \cite{Grinenko2020,Grinenko2021}, oxide interface superconductors \cite{Scheurer2015,Singh2022}, and systems exhibiting electrically or orbitally driven superconductivity \cite{Mercaldo2020,Bours2020,Mercaldo2021,DeSimoni2021,Mercaldo2023}, as well as in multiband noncentrosymmetric superconductors \cite{Hillier2009,Biswas2013,Shang2018, Singh2017,Singh2018}.
Grasping the mechanisms behind time-reversal symmetry breaking in multiband phases and pinpointing effective detection methods to explore the intricacies of multiband superconductors remains a significant challenge that has not yet been fully resolved.
A striking manifestation of the occurrence of time-reversal symmetry-breaking superconductivity is to have a zero-field superconducting diode \cite{wu22,lin2022}. 
In this context, recently, the magnetic-field free superconducting diode effect has been observed in intrinsic flakes based on kagome CsV$_3$Sb$_5$ where dynamic modulation of the polarity of the supercurrent has been ascribed as evidence of time-reversal symmetry breaking in the superconducting state \cite{Le2024}.

Focusing on Josephson devices employing conventional superconductors, the superconducting diode effect has been proposed by exploiting the combination of high harmonics in the Josephson current phase relation and applied magnetic field in single junctions or interferometric configurations \cite{souto2022,MertBorzkurt2023,MarginedaCommunPhys2023} both for dc and ac regimes \cite{Cuozzo2024} and multiterminal geometries \cite{Coraiola2024,chirolli2024,Huamani_Correa_2024}. The nonreciprocal effects with conventional superconductors have been measured in Dayem bridges \cite{MarginedaCommunPhys2023} and in interferometric setups involving superconducting quantum interference devices (SQUIDs) \cite{greco2023,greco2024}. 
%, owing to a high-harmonic content of the junctions current-phase relation (CPR).

A different perspective is given by considering multiband superconductors that can spontaneously break time-reversal symmetry.
Along this line, it has been recently shown that nonreciprocal superconducting effects can be achieved in Josephson junctions hosting multiband superconductors even without an applied magnetic field. Distinct features in the parameters space have been obtained with the multipolar distribution of the rectification amplitude, thus having a potential impact on the detection of unconventional types of pairing \cite{Yerin_diode}. 

A question now arises: Can the combined application of multiband superconductors, which undergo $\pi$-pairing or possess a superconducting order parameter that breaks time-reversal symmetry, along with a magnetic field in interferometric setups, lead to nonreciprocal effects that enhance control of the rectification amplitude and detection capabilities? 
In this paper, we face this problem by investigating the phenomena of nonreciprocal supercurrents within superconducting quantum interferometric devices (SQUIDs), which integrate Josephson junctions featuring single and multiband order parameters. Our study reveals that applying a magnetic field operating the SQUID can influence the magnitude and orientation of supercurrent rectification, in a way that depends on the multiband characteristics of the superconductors involved. 
We delve into how different pairing configurations—precisely zero and $\pi$- pairing across various bands—affect the emergence of nonreciprocal effects. Our analysis demonstrates that the presence of $\pi$-pairing does not alter the qualitative behavior of the rectification process as a function of the magnetic field. When we incorporate multiband superconductors that break time-reversal symmetry, we observe distinct signatures in the rectification behavior: the the rectification becomes even parity with respect to the magnetic field and the average rectification amplitude across different quantum flux multiples does not sum to zero. Moreover, for the case of SQUID based on combined single and three-band superconductors, the rectification amplitude is typically not vanishing at values of the magnetic flux which are half integers of the quantum flux.

\begin{figure}[h]
\includegraphics[width=0.99\linewidth]{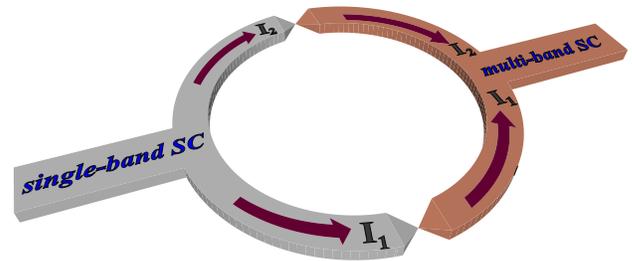}
\caption{Schematic illustration of a dc SQUID based on S-c-S type of Josephson junctions, where notation c means constriction between a single-band (grey part) and a multi-band superconductor (brown part). $I_1$ and $I_2$ define the corresponding currents through the Josephson junction.}
\label{fig:1}
\end{figure}

%\section{dc SQUID based on Josephson S-c-S type junction between a single-band and a two-band superconductor}

\noindent\large{\textbf{dc SQUID based on Josephson junctions with single- and two-band superconductors}}\normalsize\\
%with the TRSB state

%\subsection{Formulation}
\noindent
\textbf{Formulation.}
In this section, we present the formulation for deriving the expression of the supercurrents flowing within the SQUID, assuming that the junctions include a superconductor with a multiband order parameter (see Fig. \ref{fig:1}). 
We start by introducing the phase differences in the Josephson junctions between the order parameter of one band of the two-band superconductor and the order parameter of the single-band superconducting part of the SQUID loop as $\varphi _j$, where $j$ specifies the indices of the junctions ($j=1,2$). The single-valuedness of the phase differences around the loop, which are thicker than the London penetration depth for the single-band and two-band superconductors, requires \cite{Yerin_2015, Kiyko}
\begin{equation}
{\varphi _1} - {\varphi _2} = 2\pi \frac{\Phi }{{{\Phi _0}}},
\label{single_value}
\end{equation}
where $\Phi$  is the total magnetic flux and $\Phi_0$  is the flux quantum. 

It is important to note that the same quantization conditions are applicable and valid for the phase differences between the second component order parameter of the two-band superconductor and the order parameter of the single-band counterpart:
\begin{equation}
\left( {{\varphi _1} + \phi } \right) - \left( {{\varphi _2} + \phi } \right) = 2\pi \frac{\Phi }{{{\Phi _0}}},
\label{single_value2}
\end{equation}
which eventually coincides with the condition Eq. (\ref{single_value}), where $\phi$ is the intrinsic difference between the phases of the order parameters of the two-band superconductor. This phase difference $\phi$ can acquire nonzero values due to the presence of interband scattering \cite{Stanev, Corticelli, Yerin_magneto}, which is characterized by the microscopic coefficient $\Gamma_{ij} \neq 0$ (a dirty two-band superconductor). In turn, we have that when $\Gamma_{ij} = 0$ (i.e., a clean two-band superconductor), the values of $\phi$ are equal to $0$ or $\pi$ depending on the attractive or repulsive character of the interband interaction and correspond to the pairing symmetry of $s_{++}$ or $s_{\pm}$, respectively. Instead, for $\phi$ different from $0$ or $\pi$, a two-band superconductor is characterized by a complex superposition of the two order parameters that thus breaks the time-reversal symmetry. 

All calculations are performed upon the assumption that the critical temperature $T^{(s)}_{c}$ of the single-band superconducting part of the dc SQUID is at least not less than that of the multi-band counterpart $T^{(s)}_{c} \geq T^{(m)}_{c}$.

The total current in a DC SQUID can be represented as the sum of the currents flowing through the two Josephson junctions: 
\begin{equation}
I\left( \varphi_1, \varphi_2  \right) = {I_1}\left( \varphi_1  \right) + {I_2}\left( \varphi_2  \right),
\label{total_current_two}
\end{equation}
where 
\begin{widetext}
\begin{equation}
\label{current_total}
I_j(\varphi_j) = \frac{{\pi T}}{e}\sum\limits_i {\frac{1}{{{R_{Ni}}^{(j)}}}} \sum\limits_\omega  {\frac{{{C_i}}}{{\sqrt {1 - \kappa _i^2 + C_i^2} }}} \left( {\arctan \left( {\frac{{\omega {\kappa _i}{C_i} + F _i^ - \left( {{\kappa ^2} - 1} \right)}}{{\omega \sqrt {\kappa _i^2 - C_i^2 - 1} }}} \right) - \arctan \left( {\frac{{\omega {\kappa _i}{C_i}  + F _0^ - \left( {{\kappa ^2} - 1} \right) }}{{\omega \sqrt {\kappa _i^2 - C_i^2 - 1} }}} \right)} \right).
\end{equation}
\end{widetext}
Here, the Josephson phase differences $\varphi_j$ enter implicitly via the functions
\begin{equation}
\label{kappa_i}
{\kappa _i} = \frac{{F_0^ +  - F_i^ + }}{{F _0^ -   - F _i^ - }},
\end{equation}
\begin{equation}
\label{C_i}
{C_i} = \frac{{F _0^ -  F _i^ +  - F _0^ +  F _i^ - }}{{\omega \left( {F _0^ -  - F _i^ - } \right)}},
\end{equation}
where 
\begin{equation}
\label{Phi_i}
\begin{array}{l}
F _0^ \pm  = \frac{1}{2}\left( {{F _0} \pm F _0^ * } \right),\\
F _i^ \pm  = \frac{1}{2}\left( {{F _i} \pm F _i^ * } \right).
\end{array}
\end{equation}

Finally, the notation $R_{Ni}^{(j)}$ represents partial contributions to the resistance of the junctions (also known as the Sharvin resistance for the case of point contacts \cite{Sharvin}). For simplicity and to facilitate further consideration, we assume $R_{N1}^{(1)}=R_{N2}^{(1)}=R_{N3}^{(1)}=R_{N1}^{(2)}=R_{N2}^{(2)}=R_{N3}^{(2)}$ except in special cases where it will be specified.

Eq. (\ref{current_total}) is the cornerstone of our further calculations and the subsequent consideration of a dc SQUID behavior. It was derived in Ref. \onlinecite{Yerin_diode} within the Usadel equations formalism generalized for the case of multicomponent superconductivity with the interband scattering effect.

For a conventional s-wave single-band superconductor $F_0$ has simple expression:
\begin{equation}
F_0 = \left| {{\Delta _0}} \right|\exp \left( { - {\rm i}\varphi_j/2 }\right), 
\label{F0}
\end{equation}
where $\left| {{\Delta _0}} \right|$ is the modulus of the order parameter. 

For a dirty two-band superconductor, the functions $F_i$ have the following structure
\begin{equation}
\label{F1}
F_i = f_i^{\left( 0 \right)}+f_i^{\left( 1 \right)} , 
\end{equation}
where
\begin{equation}
\label{f_i0}
\begin{array}{l}
f_1^{\left( 0 \right)} = \frac{{\left( {\omega  + {\Gamma _{21}}} \right){\Delta _1} + {\Gamma _{12}}{\Delta _2}}}{{\omega \left( {\omega  + {\Gamma _{12}} + {\Gamma _{21}}} \right)}},\\
f_2^{\left( 0 \right)} = \frac{{\left( {\omega  + {\Gamma _{12}}} \right){\Delta _2} + {\Gamma _{21}}{\Delta _1}}}{{\omega \left( {\omega  + {\Gamma _{12}} + {\Gamma _{21}}} \right)}},
\end{array}
\end{equation}
and
\begin{widetext}
\begin{equation}
\label{f_i1}
\begin{array}{l}
f_1^{\left( 1 \right)} = \frac{{{\Gamma _{12}}\left( {\omega  + {\Gamma _{21}}} \right)\left( {{\Delta _1} - {\Delta _2}} \right){{\left| {f_2^{\left( 0 \right)}} \right|}^2} - \left[ {\left( {{{\left( {\omega  + {\Gamma _{21}}} \right)}^2} + {\Gamma _{12}}\left( {\omega  + 2{\Gamma _{21}}} \right)} \right){\Delta _1} + {\Gamma _{12}}\left( {\omega  + {\Gamma _{12}}} \right){\Delta _2}} \right]{{\left| {f_1^{\left( 0 \right)}} \right|}^2}}}{{\omega \left( {\omega  + {\Gamma _{12}} + {\Gamma _{21}}} \right)}},\\
f_2^{\left( 1 \right)} = \frac{{{\Gamma _{21}}\left( {\omega  + {\Gamma _{12}}} \right)\left( {{\Delta _2} - {\Delta _1}} \right){{\left| {f_1^{\left( 0 \right)}} \right|}^2} - \left[ {\left( {{{\left( {\omega  + {\Gamma _{12}}} \right)}^2} + {\Gamma _{21}}\left( {\omega  + 2{\Gamma _{12}}} \right)} \right){\Delta _2} + {\Gamma _{21}}\left( {\omega  + {\Gamma _{21}}} \right){\Delta _1}} \right]{{\left| {f_2^{\left( 0 \right)}} \right|}^2}}}{{\omega \left( {\omega  + {\Gamma _{12}} + {\Gamma _{21}}} \right)}},
\end{array}
\end{equation}
\end{widetext}
with corresponding expressions for the order parameters $\Delta_1$ and $\Delta_2$ \footnote{To avoid confusion with the notation for a magnetic flux $\Phi$, the set of $\Phi$ functions taken from the original paper Ref. \onlinecite{Yerin_diode} for the determination of the Josephson current have been redefined as $F_0$ and $F_i$.}
\begin{equation}
\label{bc_Delta1}
{\Delta _1} = \left| {{\Delta _1}} \right|\exp \left( { {\rm i}\frac{\varphi_1}{2} }\right),
\end{equation}
and
\begin{equation}
\label{bc_Delta2}
{\Delta _2} = \left| {{\Delta _2}} \right|\exp \left( { {\rm i}\frac{\varphi_2}{2} + {\rm i}\phi }\right).
\end{equation} 

Eqs. (\ref{f_i0})-(\ref{f_i1}) is valid in the region where both $\left| {{\Delta _i}} \right|$ are small, {\it i.e.} nearby the critical temperature $T^{(m)}_{c}$. The latter is determined not only by the intraband and interband interactions in a multiband superconductor but also by the interband scattering rate $\Gamma=\Gamma_{12}=\Gamma_{21}$, where the additional assumption of equal density of states at the Fermi level for each of the bands $N_1=N_2$ is introduced. This effect reduces $T^{(m)}_{c}$ compared to the critical temperature $T^{(m)}_{c0}$ of a clean two-band superconductor when $\Gamma=0$ \cite{Gurevich, Stanev}.

\begin{figure}
\includegraphics[width=1\linewidth]{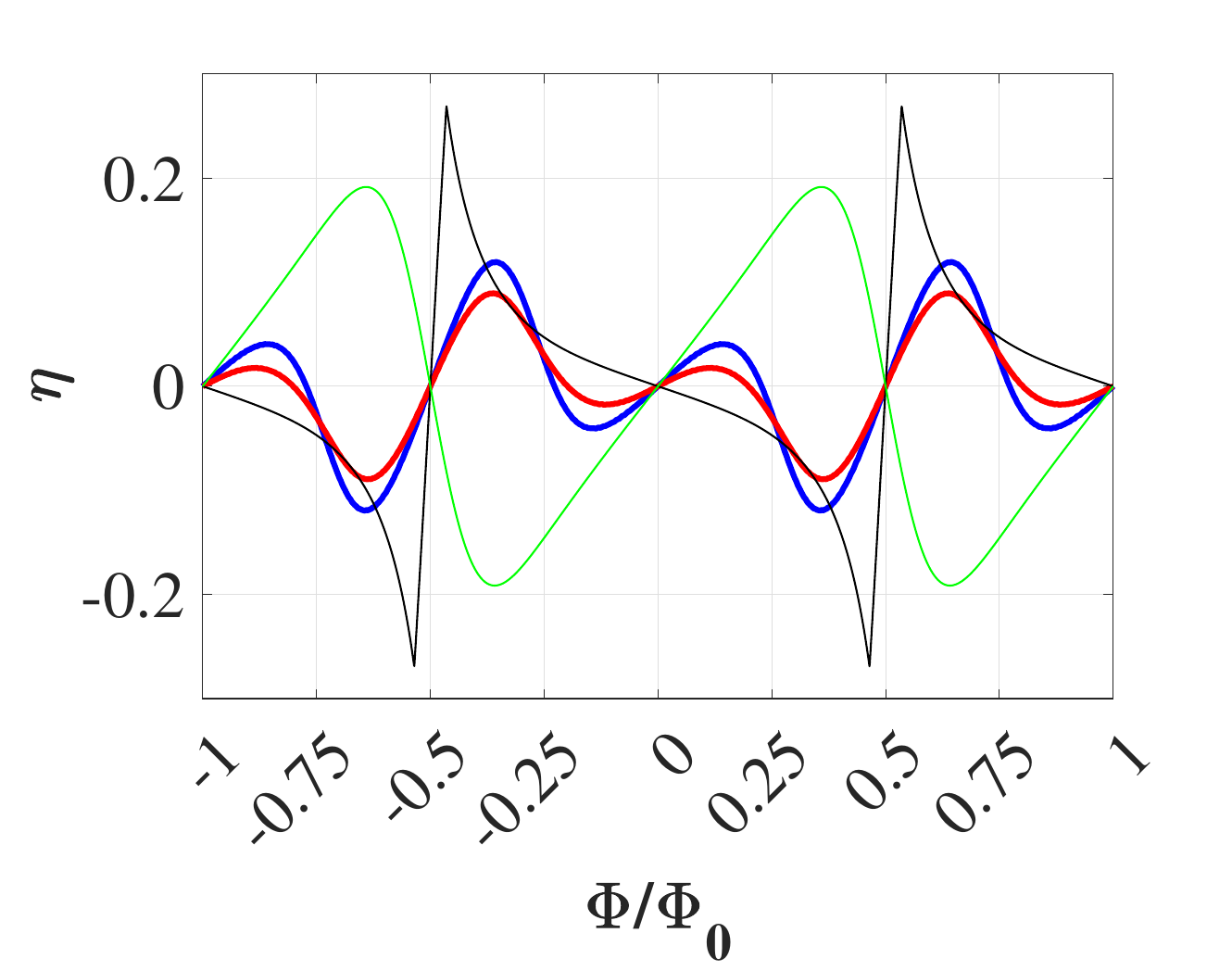}
\caption{Comparative behavior of the rectification amplitudes $\eta$ as a function of applied external magnetic flux $\Phi$ for different types of DC SQUIDs. The thin green curve illustrates the dependence of $\eta(\Phi)$ based on conventional single-band s-wave superconductors with asymmetric Josephson junction characteristics, one of which includes higher $\sin{2\varphi}$ harmonics according to Eq. (20) in Ref. \onlinecite{Fominov} with given relations between the partial contributions of the critical currents $I_{b1}/I_{a1}=2$ and $I_{b2}/I_{a1}=0.5$. The thin black curve corresponds to the dependence $\eta(\Phi)$ for dc SQUID based on S-c-S type of Josephson junctions with different transparency values $\tau_1=0.7$, $\tau_2=0.8$ and described by current-phase relations in the form of Eq. (4) in Ref. \onlinecite{Souto} (also known as a generalization of the current-phase relations of the KO-2 theory \cite{KO2, Haberkorn, Golubov_review} to the case of multichannel tunneling with different transparency \cite{Golubov_review}). The bolder blue and red curves represent the $\eta(\Phi)$ dependencies for asymmetric dc SQUID based on Josephson junctions between a conventional single-band s-wave and $s_{++}$ ($\phi=0$) and $s_{\pm}$-wave ($\phi=\pi$) dirty two-band superconductor for $\Gamma/T^{(m)}_{c0}=0.06$, respectively,  with $\left| {{\Delta _1}} \right| = 2\left| {{\Delta _0}} \right|$, $\left| {{\Delta _2}} \right| = 3\left| {{\Delta _0}} \right|$ at $T=0.7T^{(m)}_{c0}$ and with $R_{N1}^{(1)}/R_{N2}^{(1)}=1$, $R_{N1}^{(1)}/R_{N1}^{(2)}=2$ and $R_{N1}^{(1)}/R_{N2}^{(2)}=3$. }
\label{dc_SQUID_2band_simple}
\end{figure}

\textbf{The diode effect.} 
%\noindent
We start by considering the 
nonreciprocal transport in a dc SQUID, in which the two-band superconductor has no broken TRS state and is characterized by either $s_{++}$ (order parameters phase difference is $\phi=0$) or $s_{\pm}$ ($\phi=\pi$) pairing mechanism. The numerical calculation was performed using Eqs. (\ref{total_current_two}) and (\ref{current_total}) displays the absence of any sign of diode effect. This result is expected since it was shown earlier that the current-phase relation in each Josephson junction between $s$-wave and $s_{++}$ or $s_{\pm}$ two-band superconductors has odd parity.
%does not exhibit any inherent features of the junctions $\phi$ or $\varphi_0$, being solely odd functions \cite{}. 
However, when the partial contributions to the resistances of the Josephson junctions are no longer equal, this statement is violated, and the DC SQUID manifests evidence of nonreciprocal transport. In this case, according to the results of the evaluation presented in Figure \ref{dc_SQUID_2band_simple}, depending on the applied magnetic flux $\Phi$, the amplitude of rectification $\eta$ is affected by the symmetry of the order parameter in the two-band superconductor: the blue curve represents $\eta(\Phi)$ dependence for $s_{++}$ pairing mechanism and the red curve is for $s_{\pm}$ one. Interestingly, the inhomogeneity of the conducting properties of the dc SQUID allows us to distinguish two given types of the symmetry of the order parameter in the two-band superconductor exploiting the quantitative difference in the behavior of $\eta(\Phi)$.

However, such behavior may not be treated as unique, belonging solely to this type of DC SQUID. It was shown earlier that the nonmonotonic alternating character of $\eta(\Phi)$ with two peaks and zero rectification amplitude at $\Phi=\Phi_0/2$ should also occur for both an asymmetric higher harmonic SQUID, where one of the Josephson junctions has a higher harmonic of the current-phase relation \cite{Fominov}, and for a dc SQUID with different transmissions of Josephson junctions \cite{Souto} (see thin black and green curves, respectively, in Fig. \ref{dc_SQUID_2band_simple}). 

\begin{figure*}
\centering
\includegraphics[width=0.49\linewidth]{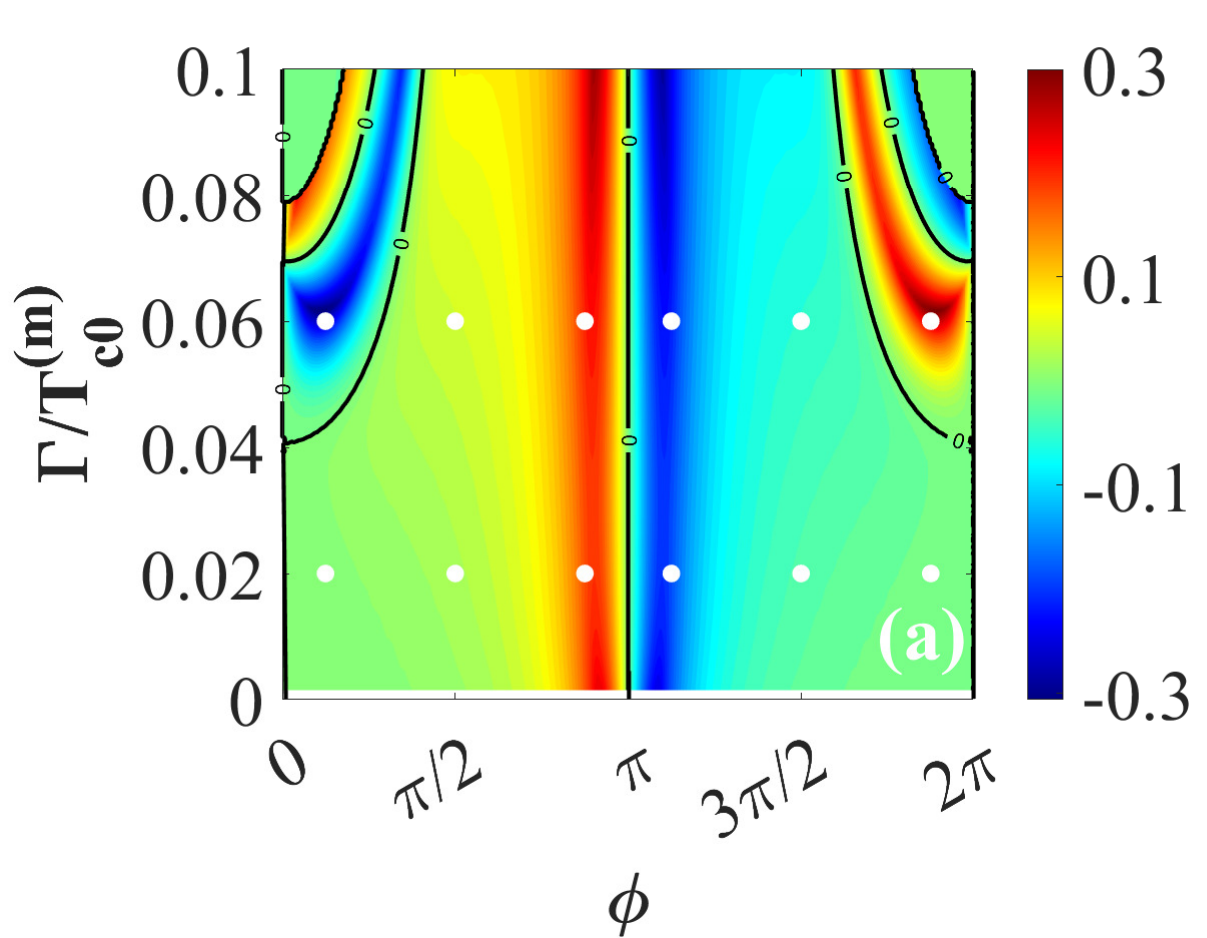}\hfil
\includegraphics[width=0.49\linewidth]{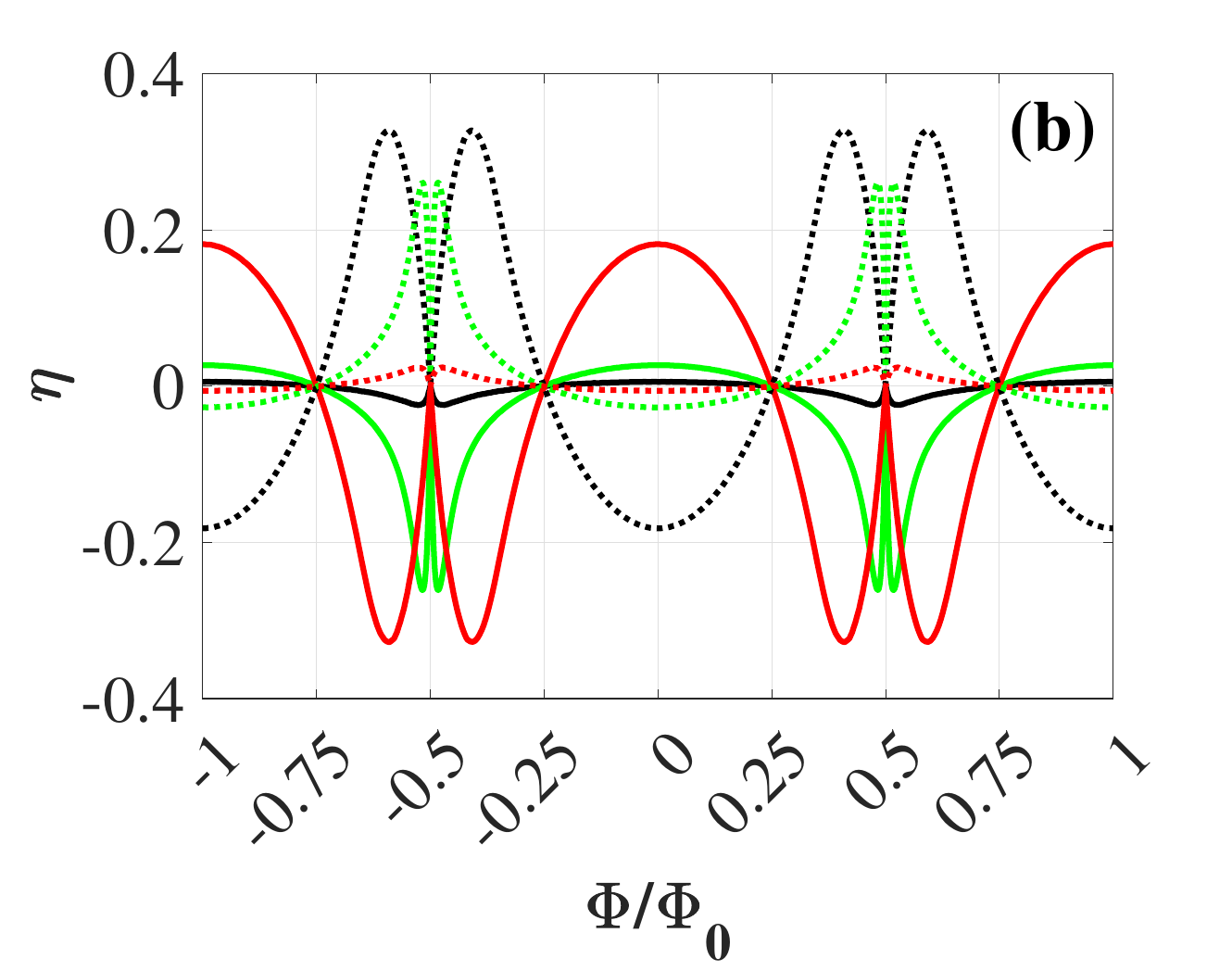}\par\medskip
\includegraphics[width=0.49\linewidth]{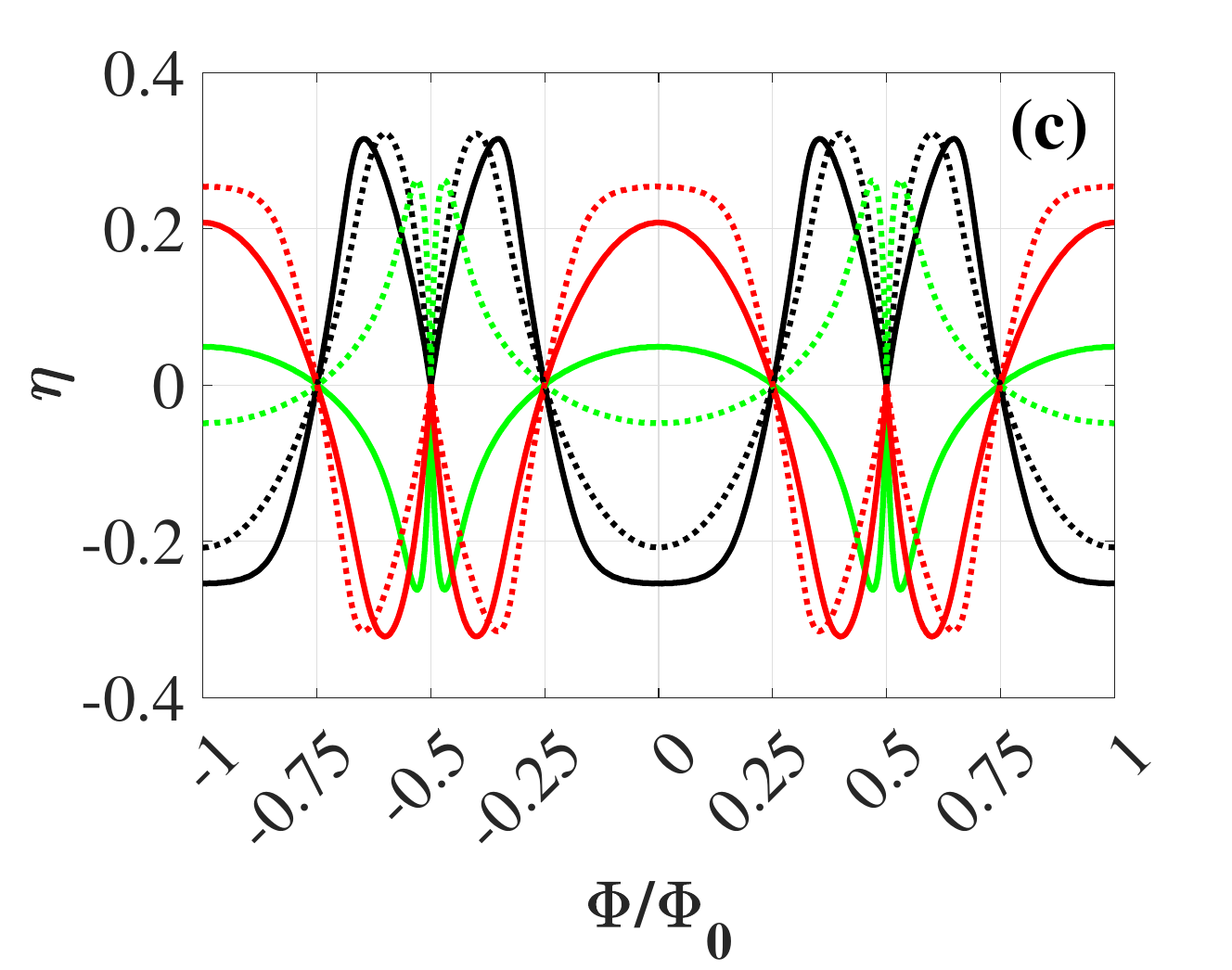}\hfil
\includegraphics[width=0.49\linewidth]{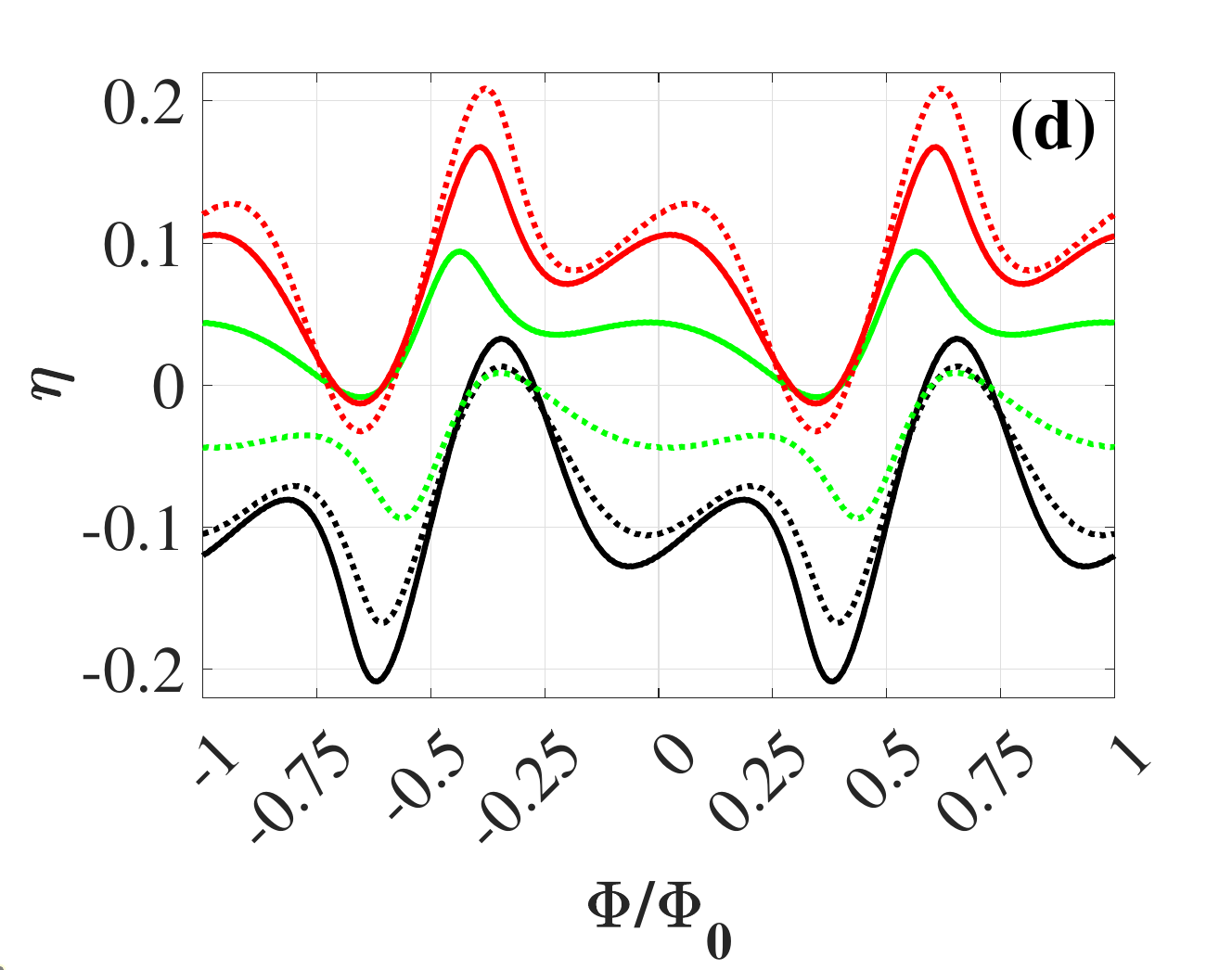}
\caption{(a) The diode rectification amplitude $\eta$ of a dc SQUID based on the Josephson junctions between a single-band and a dirty two-band superconductor as a function of the intrinsic phase difference $\phi$ and the interband scattering rate $\Gamma/T^{(m)}_{c0}$ assuming a two-band superconductor with $\left| {{\Delta _1}} \right| = 2\left| {{\Delta _0}} \right|$, $\left| {{\Delta _2}} \right| = 3\left| {{\Delta _0}} \right|$ at $T=0.7T^{(m)}_{c0}$. The white-filled circles depict the reference points for tracking the evolution of $\eta$ when an external magnetic flux is applied. (b) The diode rectification amplitude of a dc SQUID $\eta$ vs. an applied magnetic flux $\Phi/\Phi_0$ for $\Gamma/T^{(m)}_{c0}=0.02$ and $\phi=\pi/8$ (solid black), $\phi=\pi/2$ (solid green), $\phi=7\pi/8$ (solid red), $\phi=9\pi/8$ (dotted black), $\phi=3\pi/2$ (dotted green), $\phi=15\pi/8$ (dotted red). (c) The same dependencies for $\Gamma/T^{(m)}_{c0}=0.06$. (d) The same dependencies as in Figure \ref{dc_SQUID_2band}c but with the unequal contribution to the Josephson junctions resistance $R_{N1}^{(1)}/R_{N2}^{(1)}=1$, $R_{N1}^{(1)}/R_{N1}^{(2)}=2$ and $R_{N1}^{(1)}/R_{N2}^{(2)}=3$.}
\label{dc_SQUID_2band}
\end{figure*}

In the absence of an applied magnetic flux, we can construct a phase diagram of the dc SQUID as a Josephson diode showing the dependence of the rectification amplitude on the parameters of the dirty two-band superconductor: the microscopic coefficient $\Gamma$ and the intrinsic phase difference $\phi$ of the order parameters. The visualization of the numerical calculations performed within Eqs. \ref{total_current_two} and (\ref{current_total}) is presented in Figure \ref{dc_SQUID_2band}a as a contour map of function $\eta(\phi, \Gamma)$. As expected, for equal partial contributions to the resistivity of two dc SQUID Josephson junctions, it coincides with the analogous phase diagram of the Josephson junction formed by a s-wave single-band and a dirty two-band superconductor \cite{Yerin_diode}. 

To observe the evolution of the rectification amplitude of the dc SQUID when an external magnetic flux is applied, we impose a set of reference points on the constructed phase diagram with fixed values of $\Gamma$ and $\phi$ (white-filled circles in Figure \ref{dc_SQUID_2band}a) to encompass all possible hallmarks and describe the behavior of such a superconducting Josephson diode. Following these points, we plot the dependencies $\eta(\Phi)$ for $\Gamma=0.02T^{(m)}_{c0}$ (Figure \ref{dc_SQUID_2band}a), $\Gamma=0.06T^{(m)}_{c0}$ (Figure \ref{dc_SQUID_2band}b) and $\Gamma=0.08T^{(m)}_{c0}$ (Figure \ref{dc_SQUID_2band}c). For greater clarity and to better track the evolution of the rectification amplitude, we deliberately separated the curves of these functions for $\phi<\pi$ through solid lines and for $\phi>\pi$ using dashed lines. 

Now, we can outline several features of the rectification amplitude behavior.  First, it is essential to note that the diode effect is still preserved in a zero magnetic field. The diode effect vanishes when $\Phi=\Phi_0/2$ as well as when $\Phi=\Phi_0/4$ and $\Phi=3\Phi_0/4$ (Fig. \ref{dc_SQUID_2band}b and c). The last two points with $\eta=0$ result from the reciprocal compensation of the four contributions to the total current through Josephson junctions with equal partial resistances. The dependence of the rectification amplitude has a nonmonotonic sign variable character with two peaks (minima or maxima) where the coefficient $\eta$ has the same sign. Moreover, the function $\eta(\Phi)$ is even with respect to the point $\Phi=\Phi_0/2$.

%However, such similarities with our findings have origin only due to the inhomogeneous characteristics of the Josephson junctions. In fact, these dependencies of the rectification amplitude are odd functions with respect to the point $\Phi=\Phi_0/2$ instead of being even functions as in Figure \ref{dc_SQUID_2band}) when a TRSB state occurs. In addition, in the absence of applied magnetic flux, the diode effect does not occur at all in such types of a dc SQUID, i.e. $\eta(0)=0$. 

For completeness of the characterization, we additionally impose unequal contributions to the resistance of Josephson junctions and revisit the evolution of $\eta(\Phi)$ for the case of chiral symmetry of the order parameter with the specific value of $\Gamma/T^{(m)}_{c0}=0.06$. According to the numerical calculations shown in Figure \ref{dc_SQUID_2band}d, the rectification amplitude function loses its parity and acquires an asymmetric form (compare with the set of analogous dependencies in Figure \ref{dc_SQUID_2band}c)

Based on the above features of the diode effect in DC SQUID, we can introduce a quantitative criterion to identify the occurrence of the TRSB state in a multicomponent superconductor in the form of a definite integral of the function $\eta(\Phi)$:
\begin{equation}
\mathcal{I} = \int\limits_0^{{\Phi _0}} {\eta \left( \Phi  \right)} d\Phi.
\label{criterion}
\end{equation}

If $\mathcal{I}=0$, as for the rectification amplitude dependencies in Figure \ref{dc_SQUID_2band_simple}, which are odd functions, then the TRSB state does not occur in this superconductor. Otherwise, when $\mathcal{I} \ne 0$ as for $\eta(\Phi)$ shown in Figure \ref{dc_SQUID_2band}b-d, this state can emerge. It is important to note that in some exceptional cases, as shown in the following for the case of a three-band superconductor, the introduced criterion Eq. (\ref{criterion}) may fail. 
 
To study the diode effect in a dc SQUID based on a two-band superconductor, a specific intermediate temperature was chosen, which is close enough to the critical temperature of the two-band superconductor without the interband scattering effect $\Gamma=0$. Taking into account that the interband scattering suppresses the critical temperature of the two-band superconductor, such a temperature regime allows us to apply a solution of the Usadel equations for the Green functions given by the zeros of Eq. (\ref{f_i0}) and the first-order approximation as in Eq. (\ref{f_i1}). As we approach the critical temperature, we have that the first-order approximation, Eq. (\ref{f_i1}), becomes negligible, the chiral symmetry of the order parameter vanishes, and the diode effect is not realized \footnote{This occurs in the case of equal partial contributions to the Josephson junctions resistance.}. 

Conversely, with decreasing temperature, a more complicated structure of Green's functions will come into play. Consequently, the diode effect will be more pronounced, acquiring a saw-tooth-like shape of $\eta(\Phi)$ dependence similar to the thin black curve in Figure {\ref{dc_SQUID_2band_simple}}. 
\\

\noindent\large{\textbf{dc SQUID based on Josephson junctions with single- and three-band superconductors}}\normalsize\\
%with TRSB state}
%\noindent\large{\textbf{dc SQUID based on Josephson S-c-S type junction between a single-band and three-band superconductor}}\normalsize\\

\textbf{Formulation.}
The quantization condition of the phase differences of a dc SQUID can be easily generalized for the case of a three-band superconductor \cite{Kiyko, Yerin_2015}:
\begin{equation}
{\varphi _1} - {\varphi _2} = 2\pi \frac{\Phi }{{{\Phi _0}}},
\label{single_value_3band}
\end{equation}
where ${\varphi _1} - {\varphi _2}$ is the phase differences of Josephson junctions between the first order parameter of a three-band superconductor and the order parameter of the conventional s-wave single-band superconducting part of the SQUID loop. The same is true for the phase differences between the second, and third order parameters of the three-band superconductor and the order parameter of the single-band counterpart:
\begin{equation}
\left( {{\varphi _1} + \phi } \right) - \left( {{\varphi _2} + \phi } \right) = 2\pi \frac{\Phi }{{{\Phi _0}}},
\label{single_value_3band_2}
\end{equation}
and
\begin{equation}
\left( {{\varphi _1} + \theta } \right) - \left( {{\varphi _2} + \theta } \right) = 2\pi \frac{\Phi }{{{\Phi _0}}},
\label{single_value_3band_3}
\end{equation}
where the intrinsic phase differences $\phi$ and $\theta$ determine the phase differences between order parameters $\Delta_1, \Delta_2$ and $\Delta_1, \Delta_3$ of a bulk three-band superconductor. The phase differences $\phi$ and $\theta$ control the ground states of a three-band superconductor. If $\phi$ and $\theta$  are not equal to 0 or $\pi$, the ground state undergoes frustration, and the TRSB state emerges. Here, we ignore the effect of interband scattering and put $\Gamma_{ij}=0$, i.e., a dirty three-band superconductor with strong enough dominant intraband scattering compared to its interband counterparts is under consideration.

For simplicity, we consider a DC SQUID at a temperature equal to zero and with energy gaps identical to each other $| \Delta_0 |$ for both superconductors. In this case currents through junctions  $I( \varphi_1 )$ and $I\left( \varphi_2  \right)$ can be expressed as
\begin{equation}
I\left( \varphi_1, \varphi_2  \right) = {I_1}\left( \varphi_1  \right) + {I_2}\left( \varphi_2  \right),
\label{current_3band}
\end{equation}
where 
\begin{eqnarray}
\label{CPR_3band_T=0_1}
I_1\left( \varphi_1  \right) &=& \frac{{\pi \left| \Delta_0  \right|}}{{e{R_{N1}^{(1)}}}}\cos \frac{\varphi_1 }{2}{\mathop{\rm Artanh}\nolimits} [\sin \frac{\varphi_1 }{2}] \nonumber \\
 &+& \frac{{\pi \left| \Delta_0  \right|}}{{e{R_{N2}^{(1)}}}}\cos \left( \!{\frac{\varphi_1 }{2} + \phi } \!\right){\mathop{\rm Artanh}\nolimits} [\sin \left(\! {\frac{\varphi_1 }{2} + \phi }\! \right)]\nonumber \\
 &+& \! \frac{{\pi \left| \Delta_0  \right|}}{{e{R_{N3}^{(1)}}}}\cos \left(\! {\frac{\varphi_1 }{2} + \theta }\! \right){\mathop{\rm Artanh}\nolimits} [\sin \left(\! {\frac{\varphi_1 }{2} + \theta } \!\right)],
\end{eqnarray}
and
\begin{eqnarray}
\label{CPR_3band_T=0_2}
I_2\left( \varphi_2  \right) &=& \frac{{\pi \left| \Delta_0  \right|}}{{e{R_{N1}^{(2)}}}}\cos \frac{\varphi_2 }{2}{\mathop{\rm Artanh}\nolimits} [\sin \frac{\varphi_2 }{2}] \nonumber \\
 &+& \frac{{\pi \left| \Delta_0  \right|}}{{e{R_{N2}^{(2)}}}}\cos \left( \!{\frac{\varphi_2 }{2} + \phi } \!\right){\mathop{\rm Artanh}\nolimits} [\sin \left(\! {\frac{\varphi_2 }{2} + \phi }\! \right)]\nonumber \\
 &+& \! \frac{{\pi \left| \Delta_0  \right|}}{{e{R_{N3}^{(2)}}}}\cos \left(\! {\frac{\varphi_2 }{2} + \theta }\! \right){\mathop{\rm Artanh}\nolimits} [\sin \left(\! {\frac{\varphi_2 }{2} + \theta } \!\right)].
\end{eqnarray}

\begin{figure*}
\centering
\includegraphics[width=0.49\linewidth]{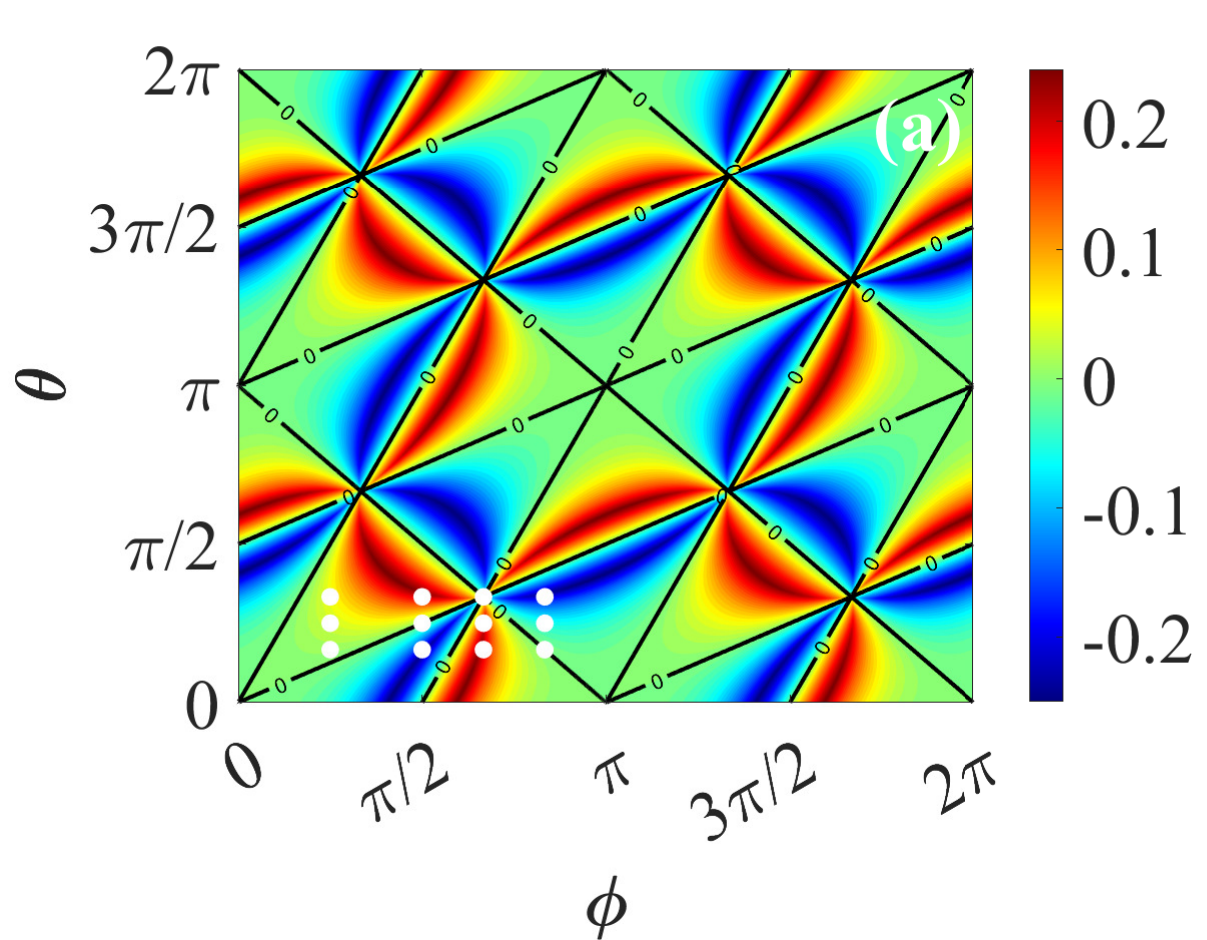}\hfil
\includegraphics[width=0.49\linewidth]{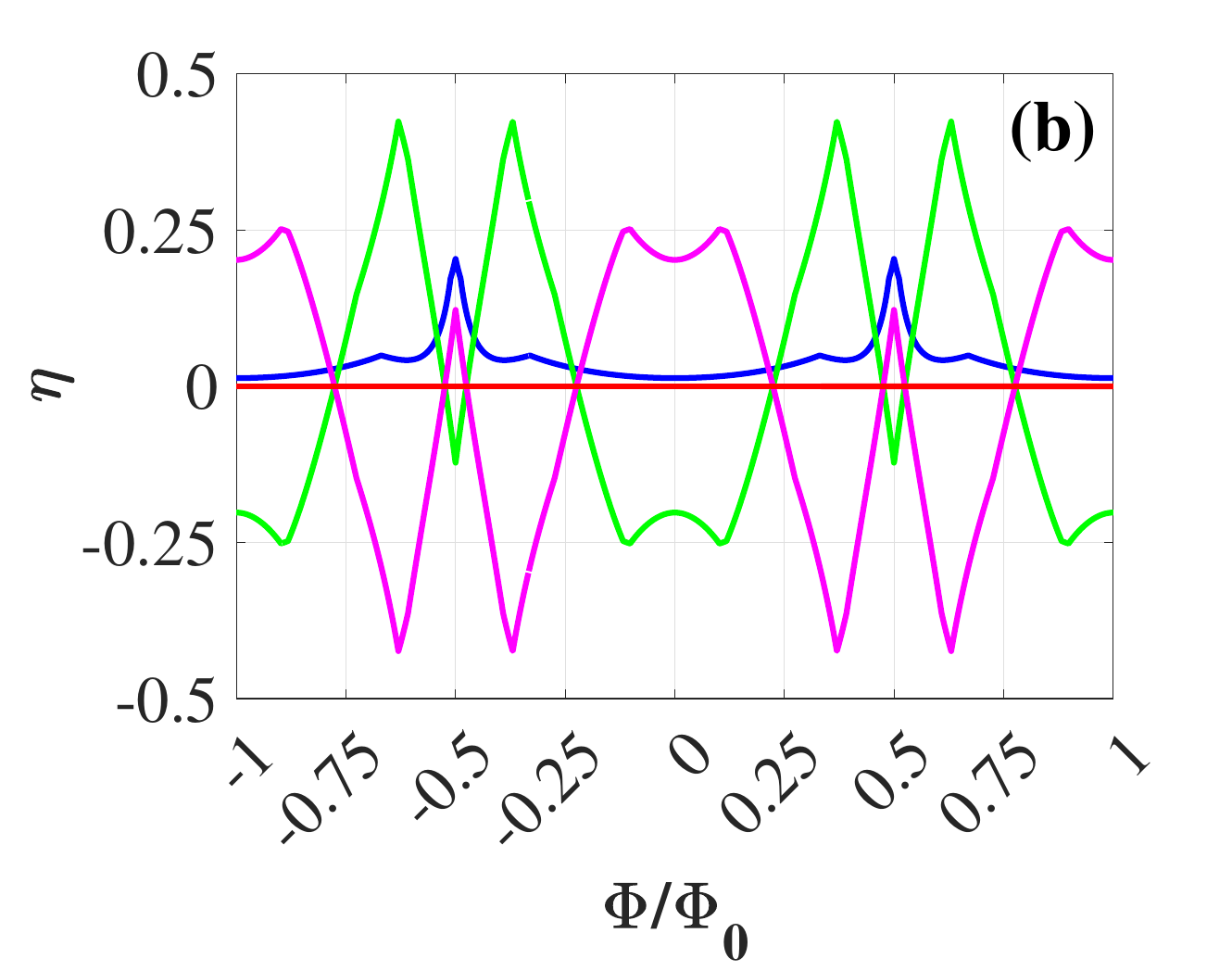}\par\medskip
\includegraphics[width=0.49\linewidth]{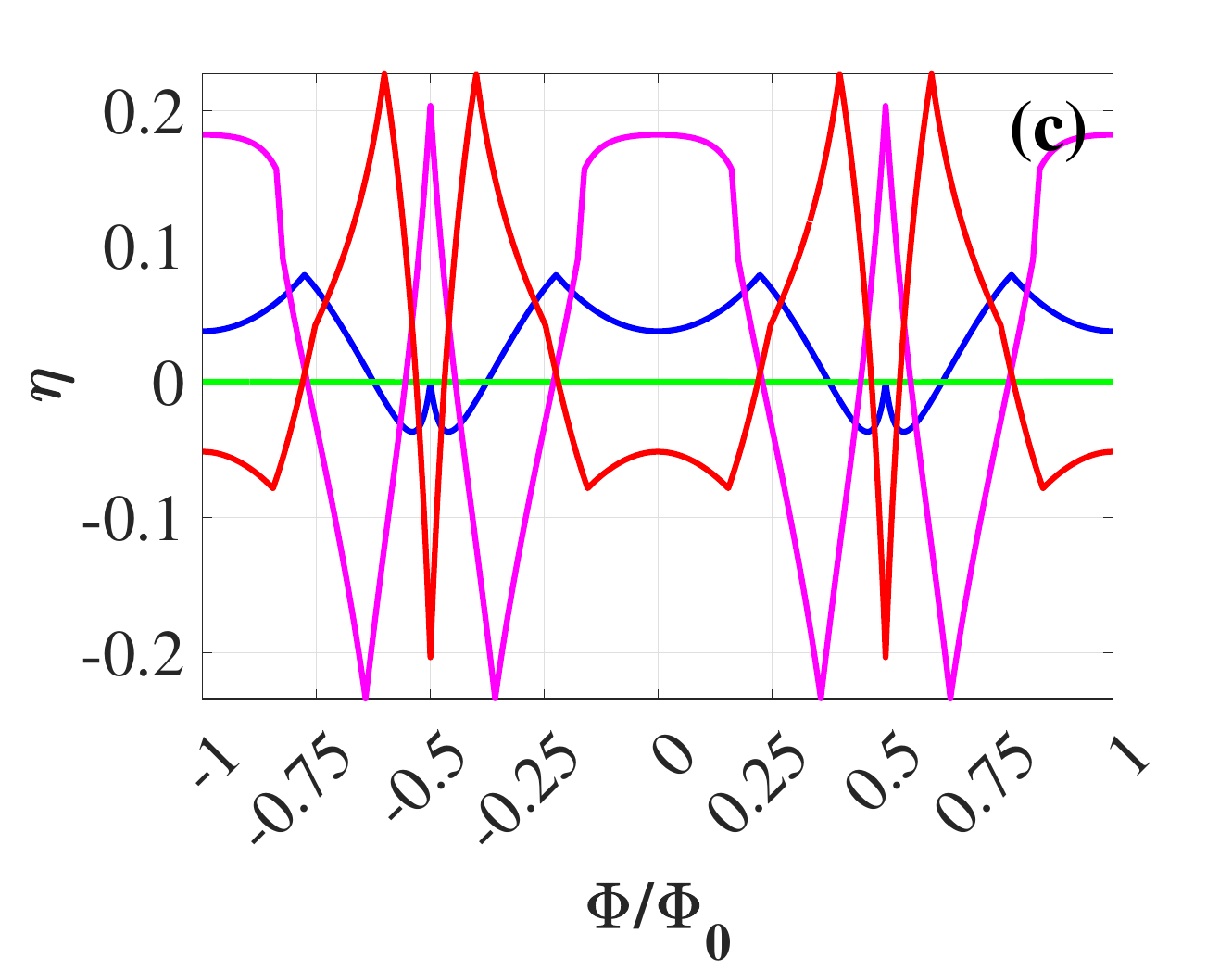}\hfil
\includegraphics[width=0.49\linewidth]{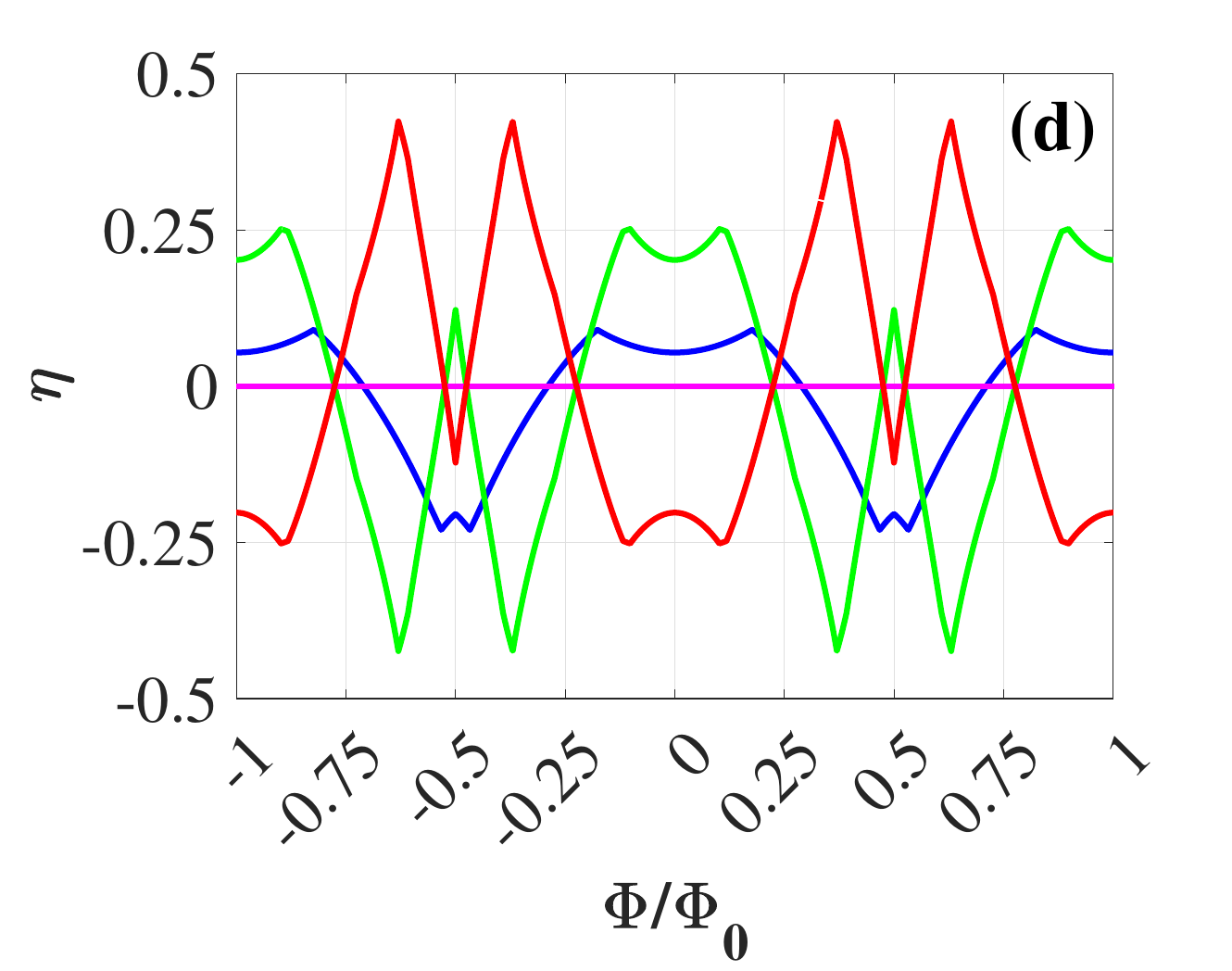}\par\medskip
\caption{(a) Diode rectification amplitude $\eta$ of a dc SQUID based on the Josephson junctions between a single-band and a three-band superconductor as a function of the internal phase differences $\phi$ and $\theta$. The filled white circles illustrate the reference points for tracking the evolution of $\eta$ when an external magnetic flux is applied (b) Dependence of the diode rectification amplitude of a dc SQUID vs. an applied magnetic flux $\Phi/\Phi_0$ for $\theta=\pi/6$ and $\phi=\pi/4$ (blue), $\phi=\pi/2$ (green), $\phi=2\pi/3$ (magenta) and $\phi=5\pi/6$ (red). (c) The same dependencies for $\theta=\pi/4$. (d) The same dependencies for $\theta=\pi/3$. }
%(e) The same dependencies for $\theta=\pi/2$. (f) The same dependencies for $\theta=2\pi/3$. (g) The same dependencies for $\theta=3\pi/4$. (h) The same dependencies for $\theta=5\pi/6$.}
\label{dc_SQUID_3band}
\end{figure*}

\textbf{The diode effect.}
Based on Eqs. (\ref{current_3band})-(\ref{CPR_3band_T=0_2}), assuming $\varphi_1=\varphi_2$ and setting as in the previous subsection equal contributions to the Josephson junctions resistance $R_{N1}^{(1)}/R_{N2}^{(1)}=1$,$R_{N1}^{(1)}/R_{N3}^{(1)}=1$, $R_{N1}^{(1)}/R_{N1}^{(2)}=1$, $R_{N1}^{(1)}/R_{N2}^{(2)}=3$ and $R_{N1}^{(1)}/R_{N3}^{(2)}=1$, one can calculate the phase diagram of the state of a Josephson diode based on dc SQUID as a function of the internal phase differences $\phi$ and $\theta$ in the absence of an external magnetic field (for comparison, we recall that in the case of a dc SQUID based on a two-band superconductor, the internal phase difference $\phi$ and the interband scattering coefficient $\Gamma$ are the governing parameters). The result presented in Figure \ref{dc_SQUID_3band}a in the form of the contour map of the function $\eta (\phi,\theta)$ predictably completely reproduces the phase diagram of a superconducting diode based on the Josephson junction between a single-band and a three-band superconductor \cite{Yerin_diode}. 

To elucidate the behavior of the DC SQUID in an external magnetic field, we use the same strategy as in the previous subsection, i.e., we use reference points (filled white circles) to study the rectification amplitude evolution as a function of the applied magnetic flux. We have deliberately arranged this set of points in the intervals $0<\phi<\pi$ and $0<\theta<\pi$ (the third quadrant of the circle centered at $\phi=\pi$ and $\theta=\pi$) because, as can be seen, the pattern exhibits properties of translational symmetry. Thus, considering Eq. (\ref{single_value_3band}) and the current-phase relations given by Eqs. (\ref{current_3band})-(\ref{CPR_3band_T=0_2}), we compute the dependencies $\eta(\Phi)$, including the so-called nodal points, where the rectification amplitude is zero at $\Phi=0$ (see Figures \ref{dc_SQUID_3band}b-d). 

%Similarly to the characterization of the peculiarities of the dc SQUID behavior based on a two-band superconductor with broken time-reversal symmetry, it is necessary to emphasize the remarkable features of this Josephson diode for the case of a three-band superconductor. 
Due to the TRSB state, the rectification amplitude $\eta$ is not equal to zero in the zero magnetic field and remarkably has an even parity upon reversal of the magnetic field. The dependence of the coefficient $\eta$ on the applied magnetic flux is qualitatively the same as for the dc SQUID based on a two-band superconductor, i.e., a sign-variable nonmonotonic function with the presence of two maxima (minima) of the same sign. Similarly, this function $\eta(\Phi)$ is symmetric with respect to the vertical line $\Phi=\Phi_0/2$. 

A hallmark of the dc SQUID with a three-band superconductor is the non-zero value of the rectification amplitude at $\Phi=\Phi_0/2$ in the vast majority of the sets $\phi$ and $\theta$ corresponding to the TRSB state. At the same time, it is notable that the nodal points remain robust against the applied magnetic flux, and the rectification amplitude is strictly zero for any value $\Phi$. For example, Figure \ref{dc_SQUID_3band} b shows that complete suppression of the rectification amplitude is achieved for $\phi=5\pi/6$ (red line) when $\theta=\pi/6$. The same occurs for $\theta=\pi/4$ (see Figure \ref{dc_SQUID_3band}c), where $\eta=0$ for $\phi=\pi/2$ (green line) and for $\theta=\pi/3$ (see Figure \ref{dc_SQUID_3band}d), when $\phi=2\pi/3$ (magenta line).

Despite the presence of nodal points where $\eta=0$ for any values of $\Phi$, in other cases, the criterion given by Eq. (\ref{criterion}) remains applicable, i.e., $\mathcal{I} \ne 0$ when the TRSB state emerges in a three-band superconductor and, correspondingly, $\mathcal{I} = 0$ if TRS is preserved in it.

%In order not to overload our consideration with figures, 
It is worth pointing out that the asymmetry in the conductive properties of Josephson junctions has the same qualitative character as in the case of DC SQUIDs based on a two-band superconductor. In other words, the functions $\eta(\Phi)$ are no longer even for a superconductor with TRSB state, evolving to an asymmetric structure with $\mathcal{I} \ne 0$ (see Figure \ref{dc_SQUID_2band_simple}b). In turn, the dependencies of $\eta(\Phi)$ for a dc SQUID based on a three-band superconductor without the TRSB state, when the internal phase differences $\phi=0, \theta=0$, $\phi=0, \theta=\pi$ or $\phi=\pi, \theta=\pi$, also demonstrate the diode effect with the same value $\mathcal{I}=0$ as in a two-band superconducting case (see Figure \ref{dc_SQUID_2band_simple}a). 

Without going into details, we also make a few remarks regarding the temperature evolution of the rectification amplitude for DC SQUID with a three-band superconductor. As the temperature increases, the Josephson current can no longer be represented as an analytical expression of Eqs. (\ref{current_3band})-(\ref{CPR_3band_T=0_2}), except near the critical temperature, where the dependence becomes sinusoidal. However, numerical calculations based on the general expression for the current of equation ((\ref{current_total})) reveal the transformation of sawtooth-like dependencies $\eta(\Phi)$ into smoother curves, and even near the critical temperature, the diode effect can be preserved with $\mathcal{I} \ne 0$ if the three-band superconductor still has a TRSB state. 
\\

\noindent\large{\textbf{Conclusions}}\normalsize\\
%\section{Conclusions} 
Having demonstrated the behavior of a SQUID in the presence of multiband superconductors, it is interesting to discuss which materials can be employed for the proposed effects.
From a practical point of view, the most suitable and promising compound for the implementation of a multicomponent superconducting element of a dc SQUID can be taken from the iron-based family, and in particular the case of ${\rm{B}}{{\rm{a}}_{\rm{x}}}{{\rm{K}}_{{\rm{1 - x}}}}{\rm{F}}{{\rm{e}}_{\rm{2}}}{\rm{A}}{{\rm{s}}_{\rm{2}}}$. As experiments have shown \cite{Grinenko2017}, at the doping level $x \approx 0.73$, a state with broken time-reversal symmetry emerges in this multi-band superconductor, with critical temperature $T^{(m)}_{c} \approx 10$ K. On the other hand, the single-band $s$-wave part of the dc SQUID can be niobium nitride NbN with $T^{(s)}_{c} \approx 16$ K or some compounds of the superconducting A15 family (such as ${{\rm{V}}_{\rm{3}}}{\rm{Si}}$ with $T^{(s)}_{c} \approx 16$ K) or even a two-band superconducting magnesium diboride ${\rm{Mg}}{{\rm{B}}_{\rm{2}}}$ with $T^{(s)}_{c} \approx 39$ K, in which both components of the order parameter have an isotropic $s$ wave structure without the presence of the broken time-reversal state.

Other relevant materials to be employed are those belonging to the class of kagome materials. In this context, our results may account for the qualitative features observed in the superconducting field-free diode effects for the CsV$_3$Sb$_5$ kagome system \cite{Le2024}. We expect that interference between domains with different phases among the bands dependent order parameters will lead to asymmetric rectification amplitude in the applied magnetic field. Additionally, we argue that thermal cycles can lead to slight changes in the phase of the superconducting order parameter, which would be associated with a sizable modification of the polarity of the nonreciprocal response. This is explicitly demonstrated by our results when comparing time-reversal broken configurations.

Regarding the hallmarks of the supercurrent rectification interferometric setup, we point out those related to the configurations integrating multiband superconductors with broken time reversal symmetry. Our findings show that the dependence on the magnetic field becomes even parity for junctions having equal resistance or they do not have any parity for generic configurations. Hence, there is a net and sizable rectification amplitude when sweeping the magnetic field over positive and negative values or over multiple quantum fluxes. Then, due to the nonvanishing rectification even in the presence of magnetic field variation, we argue that with these types of SQUIDs one can envision the design of a magnetic flux pump or magnetic field rectifier.

%\section{Conclusions}

\noindent\large{\textbf{Data availability}}\normalsize\\
\noindent {The data that support the findings of this study are available
from the corresponding author upon reasonable request.}

\bibliography{REF}

%\begin{acknowledgments}
\noindent\large{\textbf{Acknowledgments}}\normalsize\\ 
F.G. and M.C. acknowledge the EU’s Horizon 2020 Research and Innovation Framework Programme under Grants No. 964398 (SUPERGATE),  and the PNRR MUR project PE0000023-NQSTI for partial financial support. F.G. also acknowledges the EU’s Horizon 2020 Research and Innovation Framework Programme under Grants No. 101057977 (SPECTRUM).
%\end{acknowledgments}

\noindent\large{\textbf{Author contributions}}\normalsize\\

\noindent Y.Y., M.C. and F.G. conceived the project. Y.Y. performed the computations. Y.Y., M.C., F.G., S.-L.D., and A.A.V. analysed and discussed the results and the implications equally at all stages. The manuscript has been written by Y.Y, M.C. and F.G. with the help of all the authors.

\noindent\large{\textbf{Competing interests}}\normalsize\\
The authors declare no competing interests.\\

\noindent\textbf{Correspondence} and requests for materials should be addressed to Yuriy Yerin.

\end{document}